\documentclass[aps,prl,twocolumn,superscriptaddress,showpacs]{revtex4-1}
\usepackage{epsfig}
\usepackage{amsmath}
\usepackage{graphicx}
\usepackage{alltt}
\usepackage{latexsym}
\newcommand{\be}{\begin{equation}}
\newcommand{\ee}{\end{equation}}
\newcommand{\benn}{\begin{displaymath}}
\newcommand{\eenn}{\end{displaymath}}
\newcommand{\ba}{\begin{eqnarray}}
\newcommand{\ea}{\end{eqnarray}}

\renewcommand{\vec}[1]{\mbox{\boldmath $#1$}}

\renewcommand{\d}[1]{{\rm d}#1}

\newcommand{\Rbs}{Rayleigh-Brillouin scattering}

\newcommand{\srbs}{spontaneous Rayleigh-Brillouin scattering}

\begin{document}
\title{Rayleigh-Brillouin Scattering in Binary Gas Mixtures}
\author{Z. Gu}
\affiliation{Department of Physics and Astronomy, and LaserLab, VU University, De Boelelaan 1081,
1081 HV Amsterdam, The Netherlands}
\author{W. Ubachs}
\affiliation{Department of Physics and Astronomy, and LaserLab, VU University, De Boelelaan 1081,
1081 HV Amsterdam, The Netherlands}
\author{W. Marques Jr}
\affiliation{Departamento de F\'{\i}sica, Universidade Federal do
Paran\'a, Caixa Postal 10944, 81531-990, Curitiba, Brazil}
\author{W. van de Water}
\affiliation{Physics Department, Eindhoven University of Technology,
Postbus 513, 5600 MB Eindhoven, The Netherlands}
%
\date{\today}

\begin{abstract}
\noindent
Precise measurements are performed on spectral lineshapes of \srbs\
in mixtures of the noble gases Ar and Kr, with He.  Admixture of a light
He atomic fraction results in marked changes of the spectra, although
in all experiments He is merely a spectator atom: it affects the
relaxation of density fluctuations of the heavy constituent, but its
contribution to the scattered light intensity is negligibly small.
The results are compared to a theory for the spectral lineshape
without adjustable parameters, yielding excellent agreement for the case of binary
mono-atomic gases, signifying a step towards modeling and understanding of light scattering
in more complex molecular media.

\end{abstract}


\maketitle

The spectrum of light scattered in a gas is determined by the
fluctuations of its refractive index~\cite{Strutt1899}, or,
equivalently, by the motion of its molecules.  When the mean free
path between collisions is much larger than the wavelength, the
scattering spectral lineshape is a pure Gaussian, to be understood as a Doppler
effect.
At higher pressures collisional excitations and acoustic modes come
into play, as was recognized independently by
Brillouin~\cite{Brillouin1922} and Mandelstam~\cite{Mandelstam1926}.
In first approximation, redshifted and blue-shifted frequency
components are added to the scattering spectrum with characteristic
shifts $\Delta \nu = v_s \: k / 2\pi$, with $v_s$ the speed of sound
and $k$ the size of the scattering wavevector, $k / 2\pi = 2
\sin(\theta/2)/\lambda$ with $\theta$ the scattering angle and
$\lambda$ the wavelength of the incident light.

\Rbs\ in dilute gases offers a sensitive probe of gas kinetics.
Understanding the scattered light spectrum involves the linearized
Boltzmann equation \cite{Leeuwen1965}, and throughout the years
intricate approximations to the collision integral have resulted in
various kinetic models for the scattered light spectrum.  These
models may be viewed as a success of statistical physics.  Still,
discrepancies with experiments exist, and the kinetic models are
generally restricted to simple gases.
In contrast, the Earth's atmosphere consists of a mixture of gases, each of which explores
internal molecular degrees of freedom.
An important practical application of understanding such mixtures of gases is in its connection to laser light scattering
(LIDAR) of the atmosphere \cite{She1992,Witschas2014b}, in particular the
ADM-Aeolus mission of the European Space Agency
for measuring the global wind profile~\cite{Reitebuch2009}.

The Tenti model is a well-known theory for the spectral lineshape of
scattered light in monomolecular gases \cite{Boley1972,Tenti1974}.
The spectrum is determined by the communication between kinetic and
internal degrees of freedom, which is characterized by a transport
coefficient, the bulk viscosity $\eta_b$.  The bulk viscosity is a
dynamic quantity, which is not well known at the GHz frequencies of
interest in light scattering.
Therefore, $\eta_b$ was used as an adjustable parameter to describe
spectral profiles in light scattering in both
coherent~\cite{Grinstead2000,Pan2002,Pan2005,Meijer2010} and
spontaneous~\cite{Vieitez2010,Gu2013b,Gu2014a} Rayleigh-Brillouin
scattering experiments.
However, the Tenti model is not designed to describe light scattering in
mixtures. Nevertheless, applying it to air and assuming that air is a
fictitious gas with effective values for its transport coefficients
and molecular mass, yields fair agreement with
experiments~\cite{Gu2013a,Gu2014b}.
When devising a proper theory for air, one faces the formidable task
of including both kinetic and internal degrees of freedom for several
species.

As a first step in understanding light scattering in more complex gases, we will
concentrate on mixtures of noble gases using a new experimental setup
which provides spectra with unprecedented statistical accuracy
\cite{Gu2012rsi}.
We will compare these spectra to models with no adjustable
parameters; the only parameter needed is the atomic diameter, which
follows from the well-known value of the shear viscosity of the pure
noble gas \cite{Chapman1970}.  All our experiments are in the kinetic
regime, where the mean-free path between collisions is comparable to
the scattered light wavelength. Interestingly, mixtures of gases with
very different mass behave in a similar fashion as a gas of molecules
with internal degrees of freedom. While the two components of the
mixture briefly can have different temperatures, a molecular gas can
have different temperatures associated with translational and
internal degrees of freedom. It is the relaxation of these
temperature differences that determines the scattered line shape.

\begin{figure}
\includegraphics[width=1.\columnwidth]{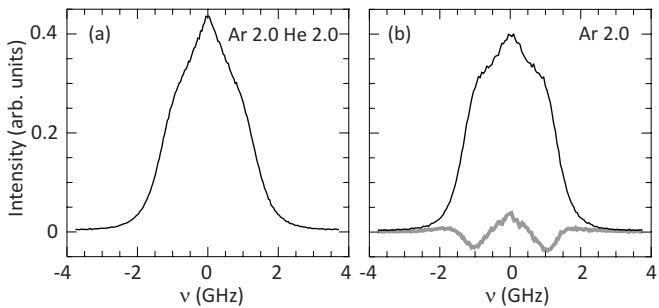}
\caption{
   In He-Ar mixtures, the light He atom acts as a {\em spectator}, it
   participates in the gas kinetics, but does not contribute to the
   scattered light intensity.  The spectra of a mixture of 2 bar Ar
   and 2 bar He (a) is very different from a spectrum of 2 bar Ar
   (b).  The gray line indicates the difference between the spectra.}
\label{fig.res.diff}
\end{figure}

The components of our He-Ar and He-Kr mixtures have a large mass
disparity ($M_{\rm He} / M_{\rm Ar} = 0.1002$, and $M_{\rm He} /
M_{\rm Kr} = 0.0478$.  With these different size atoms, it is only
the heavy ones that contribute to the scattered light intensity.
The intensity is proportional to the square of the optical
polarizabilities $\alpha$, with the ratio $(\alpha_{\rm He} /
\alpha_{\rm Ar})^2 = 1.56\times10^{-2}$, and $(\alpha_{\rm He} /
\alpha_{\rm Kr})^2 = 5.96\times10^{-3}$.
Thus, the light atoms are {\em spectators}, and influence the
spectral line shape only indirectly through collisions. Nevertheless,
as Fig.\ \ref{fig.res.diff} illustrates, their influence can be
large: adding light atoms to a gas of heavy ones significantly
changes the shape of the scattered light spectrum.

\begin{figure}
\includegraphics[width=1.\columnwidth]{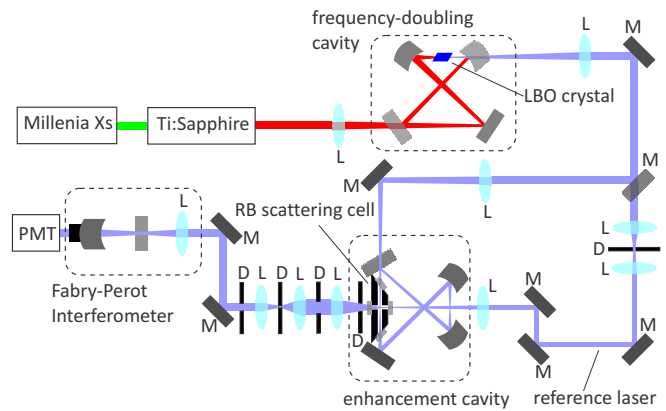}
\caption{(Color online) Schematic diagram of the experimental setup for
   \srbs. The laser beam (blue line) is amplified in an enhancement
   cavity to increase the scattering intensity. Scattered light at an
   angle of $90^\circ$ is collimated and directed onto a
   piezo-scannable Fabry--Perot interferometer for spectral analyses,
   and detected on a photomultiplier tube (PMT).}
\label{fig.s.setup}
\end{figure}

A schematic view of the setup for the measurement of \srbs\ is shown
in Fig.\ \ref{fig.s.setup}.  The light from a narrowband
continuous--wave laser is scattered off a gas contained in a
temperature--controlled gas cell.
The laser is a frequency--doubled Ti:Sa laser delivering light at
403~nm, 2~MHz bandwidth and 400~mW of output power.  The long--term
frequency drift was measured with a wavelength meter to be smaller
than 10~MHz per hour.
The scattered light is collected at an angle of $90^\circ$ from an
auxiliary focus inside the enhancement cavity, in which a
scattering--cell is mounted. The cell is sealed with Brewster
windows. The enhancement cavity amplifies the circulating power
delivering a scattering intensity of 4 Watt in the interaction region
\cite{Gu2012rsi}. The light that passes through the FPI is detected
using a photo multiplier tube (PMT) which is operated in the
photon-counting mode and read out by computer. All measurements are
performed at room temperature, $297 \pm 1$ K.

The scattering angle is determined to be $90 \pm 0.9^\circ$ by means
of the reference laser beam and geometrical relations using sets of
diaphragms and pinholes present in the optical setup.  The scattered
light is filtered by a diaphragm which covers an opening angle of
$2^\circ$, collected by a set of lenses, further filtered by an extra
pinhole ($d = 50 \: \mu{\rm m}$) and then directed into a
hemispherical scanning Fabry--Perot interferometer, which is used to
resolve the frequency spectrum of the scattered light. To scan the
FPI plate distance, the spherical mirror is mounted on a
piezo--electrical translator, which is controlled by computer.

The spectral response $S(\nu)$ of the Fabry--Perot spectrometer was
measured in a separate experiment, and could be parametrized very
well by the formula
$S(\nu) = \left[ 1 + 4 (\nu / \nu_{\rm w})^2 {\rm sinc}^2(\pi \nu/
\nu_{\rm FSR}) \right]^{-1}$, with ${\rm sinc}(x) = \sin(x)/x$, and
where $\nu_{\rm FSR}$ is the free spectral range of the etalon,
$\nu_{\rm FSR} = 7553$~MHz, and $\nu_{\rm w} = 139$~MHz is the
Airy--width of the transmission peak. All computed model spectra were
convolved with $S(\nu)$, and since the free spectral range is
relatively small, it is important to allow for the periodic nature of
$S(\nu)$.

The light scattering experiments do not provide an absolute
intensity, therefore the experimental and computed spectra were
normalized such that $\int_{-\nu_b}^{\nu_b} I(\nu) \: \d \nu = 1$,
where the integral extends over one free spectral range (FSR), $\nu_b
= \nu_{\rm FSR} / 2$.
Assuming Poissonian statistics of registered photon counts, an
estimate of the statistical error $\sigma(\nu_i)$ of measured spectra
was obtained from the square root of the accumulated photon count
$N_i$ at each discrete frequency $\nu_i$.  It was verified that the
fluctuations $N_i^{1/2}$ at each $\nu_i$ were independent.
The normalized error is then
$\chi^2 = N^{-1} \sum_{i=1}^N [I_m(\nu_i) - I_e(\nu_i)]^2 /
\sigma^2(\nu_i)$.
If the computed line shape model $I_m$ would fit the measurement
perfectly, then only statistical errors remain and the minimum of
$\chi^2$ is unity.  The difference between theory and experiment will
be expressed by $\chi^2$.

%
In the past decades, many ingenious efforts have been undertaken to
arrive at approximate solutions of the Boltzmann equation which are
relevant for light scattering.
Light scattering involves density fluctuations, with the spectrum of
scattered light equalling the Fourier transform of the density-density
correlation function.  Van Leeuwen and Yip showed that this
correlation function follows from the first moment of the solution of
the linearized Boltzmann equation~\cite{Leeuwen1965}.

\begin{figure}[t]
\includegraphics[width=1.\columnwidth]{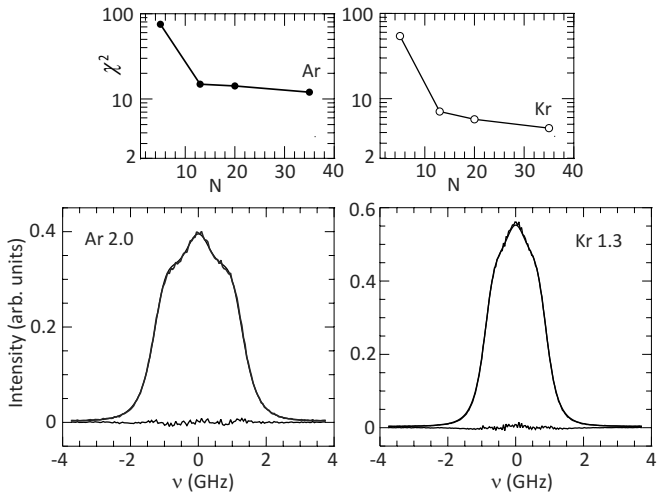}
\caption{
   Comparison of Ar at $p = 2 \; {\rm bar}$ and Kr at $p = 1.3 \;
   {\rm bar}$ spectra to the $N-$moment model for $N = 5, 13, 20$
   and 35.  The spectra are shown for $N = 35$, while the convergence
   with increasing number of moments $N$ is demonstrated in the top panels.
   }
\label{fig.res1}
\end{figure}

One such effort is based on the Bhatnagar-Gross-Krook (BGK) model,
which takes a simple relaxation form for the collision
integral~\cite{Bhatnagar1954},
\be
   \frac{\partial f}{\partial t} + (\vec{c}\cdot\nabla) f =
   -\sigma\: (f - f_r),
\label{eq.bgk}
\ee
with $\vec{c}$ the molecular velocity, $f$ the position-velocity
distribution function, and $f_r$ a reference distribution function.
The latter is determined from the requirement that $N$ of its moments
are the same as those of the complete collision integral for
monatomic particles with a $r^{-4}$ repulsive interaction potential
\cite{Marques1998}. Through increasing $N$, increasingly accurate
predictions of light scattering spectra can be computed.  We show
this convergence for light scattering in pure samples of Ar and Kr in
Fig.\ \ref{fig.res1}.
These experiments are in the kinetic regime, with uniformity
parameter $y$ of order one.  The uniformity parameter $y$ of a simple
gas is the ratio of the scattering wavelength to the mean free path
$l$ between collisions, $y = 1 / (k \: l)$.
The only information that the theory further needs is the hard-sphere
diameter $a$ of the noble gas atom, $a_{\rm Ar} = 3.66\times 10^{-10}
\: {\rm m}$ and $a_{\rm Kr} = 4.20\times 10^{-10} \: {\rm m}$.
In the case of Ar, the uniformity parameter is $y_{\rm Ar} =
   1.14$, while for Kr $y_{\rm Kr} = 0.96$.
The $N = 35$ model can hardly be distinguished from the experimental data.
However, since the minimum $\chi^2$ is still larger than 1, a slight
but significant difference between model and experiment remains,
although on a relative scale it is $\lesssim 1$\%.
In fact, the agreement is so good that the model might be viewed as a
benchmark testing for experiments.
%
%
It should be realized, however, that a monatomic ideal noble gas is the
simplest system thinkable as there are no internal molecular degrees
of freedom.

%
The BGK-moment approach was used to develop a theory of light
scattering in binary mixtures of noble gases~\cite{Bonatto2005}.
There, the reference distribution function $f_r$ in Eq.\
(\ref{eq.bgk}) was chosen such as to satisfy the principal
conservation properties of the full Boltzmann collision operators,
whilst requiring complete correspondence with two-fluid
hydrodynamics, i.e. the generalized equations of Navier-Stokes and
Fourier.  Therefore this model provides a precise transition between
the hydrodynamic and kinetic regimes \cite{Fernandes2004}.
In this sense, the BGK model with a judicious selection of the
reference distribution $f_r$ allows one to include the relevant
physical phenomena in the kinetic description. This is a great
advantage for the design of models, as the computation of these
various contributions is cumbersome.
It is expected that a model that works well for mixtures, no longer
works if the density of one of the constituents vanishes, and the gas
becomes monatomic.  This is because in the design of the mixture
model the focus is on inter-species relaxation of temperatures and
velocities, and not on the relaxation of high-order gradients that
determine the shape of the spectrum of light scattered from monatomic
gases.

\begin{figure}[t]
\includegraphics[width=1.\columnwidth]{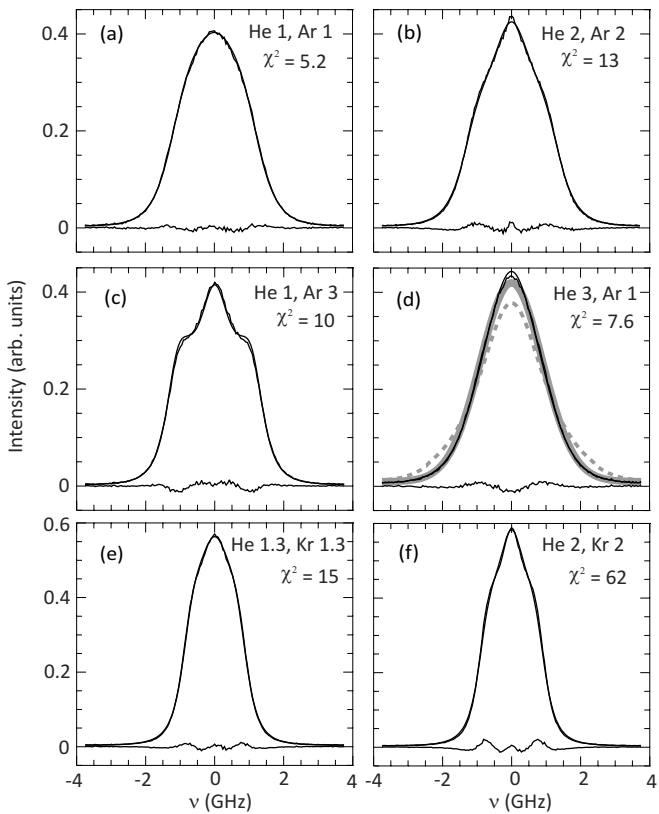}
\caption{
   (Color online) Comparing scattered light spectra of mixtures to the mixture model
   of~\cite{Bonatto2005}.
(a) Equimolar mixture He $p = 1\; {\rm bar}$, Ar $p = 1\; {\rm bar}$,
uniformity parameters $y_{\rm Ar} = 1.63, y_{\rm He} = 0.54$,
(b) equimolar mixture He $p = 2\; {\rm bar}$, Ar $p = 2\; {\rm bar}$,
$y_{\rm Ar} = 3.28, y_{\rm He} = 1.08$,
(c) asymmetric mixture He $p = 1\; {\rm bar}$, Ar $p = 3\; {\rm bar}$,
$y_{\rm Ar} = 2.96, y_{\rm He} = 1.16$,
(d) asymmetric mixture He $p = 3\; {\rm bar}$, Ar $p = 1\; {\rm bar}$,
$y_{\rm Ar} = 3.59, y_{\rm He} = 1.00$,
(e) equimolar mixture He $p = 1.3\; {\rm bar}$, Kr $p = 1.3\; {\rm bar}$,
$y_{\rm Kr} = 3.28, y_{\rm He} = 0.77$,
(f) equimolar mixture He $p = 2\; {\rm bar}$, Kr $p = 2\; {\rm bar}$,
$y_{\rm Kr} = 5.01, y_{\rm He} = 1.17$.
   The lower line is the difference between experiment and model.
   Apart from the normalization of the spectra there are no
   adjustable parameters.  In (d), the dashed thick (gray) line shows the
   Lorentzian spectral lineshape $I_D$ Eq.\ (\ref{eq.dicke}) with
   $D_{\rm He-Ar}$ the hard-sphere value, $D_{\rm He-Ar} = 1.52\times
   10^{-5}\;{\rm m}^2 {\rm s}^{-1}$.
   The full thick (gray) line (partly obscured) shows the purely Gaussian
   spectrum of Ar.}
\label{fig.res2}
\end{figure}

Measured and computed spectra for light scattering in He-Ar and He-Kr
mixtures are shown in Fig.\ \ref{fig.res2}. They are characterized by
the (partial) pressures and the (partial) uniformity parameters.
The definition of the partial uniformity parameters $y_i = 1 / (k
l_i)$ of a binary mixture involves the partial mean free paths $l_i$
of hard-sphere atoms,
\be
   l_i = \left(\pi \sum_{j = 1}^2 n_j a_{i j}^2
   \: \sqrt{1 + M_i / M_j}\right)^{-1},
\label{eq.y}
\ee
with $n_i$ and $M_i$ the number density and atomic mass of
constituent $i$, respectively, and $a_{i j} = (a_i + a_j) / 2$ the
distance between the centers of two spherical particles with
diameters $a_i$ and $a_j$ at the instant of collision.
With all uniformity parameters of order one, the experiments are in
the kinetic regime.
The computation of the theory\ \cite{Bonatto2005} now also needs the
hard-sphere diameter of He, $a_{\rm He} = 2.16\times 10^{-10} \: {\rm
m}$, and the atomic polarizabilities of the noble gases
$\alpha_{\rm He} =
0.227 \times 10^{-40}\; {\rm C}{\rm m}^2\: {\rm V}^{-1}$,
$\alpha_{\rm Ar} =
1.82 \times 10^{-40}\; {\rm C}{\rm m}^2\: {\rm V}^{-1}$,
$\alpha_{\rm Kr} =
2.94 \times 10^{-40}\; {\rm C}{\rm m}^2\: {\rm V}^{-1}$.

The measured mixture spectra are reproduced well by the theory.
Although the mixture model is designed to represent the relevant
interspecies relaxation processes, which become less important for
asymmetric mixtures, the agreement with the measured spectra at 1 bar
He and 3 bars Ar, and the reverse case, is still excellent.
%

Early experiments on mixtures of He and Xe atoms were done by
Clark~\cite{Clark1975}, but were interpreted in terms of
hydrodynamics, as a complete kinetic theory was still lacking.
Light scattering on He-Xe mixtures in a range of pressures comparable
to ours was studied by Letamendia {\it et al.}~\cite{Letamendia1982},
and sizable differences with a kinetic mixture
model~\cite{Boley1972b} were found.

A striking observation is the narrowing of the mixture spectrum in
Fig.\ \ref{fig.res2}(a,d) when the number of helium atoms is
increased while keeping the number of Ar atoms the same.  Collisional
narrowing takes place when for Ar, collisions with He atoms become
far more numerous than collisions with another Ar atom.  It stands
out in the strongly asymmetric mixture of Fig.\ \ref{fig.res2}(d).
In the kinetic case ($y_{\rm Ar} = {\cal O}(1)$), the spectral
lineshape is a Gaussian, $I(f) \propto \exp(-(2\pi \nu / k v_0)^2)$,
with $v_0$ the Ar thermal velocity, while in the hydrodynamic limit
($y_{\rm Ar} \gtrsim 10$), the spectral lineshape is a Lorentzian:
the Fourier transform of the fundamental solution of the diffusion
equation,
\be
   I_D(\nu) = \frac{2 D_{\rm He-Ar} k^2}{ (2\pi \nu)^2 + (D_{\rm
   He-Ar} k^2)^2},
\label{eq.dicke}
\ee
with $D_{\rm He-Ar}$ the mutual He-Ar diffusion coefficient.
This collisional narrowing is known as Dicke
narrowing~\cite{Dicke1953}; in the hydrodynamic limit it can be
explained by the decoherence collisions which an atomic scatterer
experiences when it straggles diffusively through a dense gas of
spectators.
Remarkably, spectral narrowing is manifested here in {\em
non-resonant} light scattering.
In our case the partial uniformity parameter evolves from $y_{\rm Ar}
= 1.6$ at $p_{\rm He} = 1 \; {\rm bar}$ to $y_{\rm Ar} = 3.6$ at
$p_{\rm He} = 3 \; {\rm bar}$.  Figure\ \ref{fig.res2}(d) illustrates
that our experimental conditions are still far from the hydrodynamic
limit, as the experiment is closer to the Gaussian than to the
Lorentzian lineshape.  By design, our theory for the lineshape
embodies both extreme cases, and it provides a near perfect
reproduction of the experimental spectrum.

We have studied Rayleigh-Brillouin scattering in mixtures of noble
gases, whose constituents have a very different mass.  In all cases,
the addition of the light He atomic gas has a large influence on the
spectral line shapes, although He atoms hardly contribute to the
scattered light intensity.
Density fluctuations in mixtures are dominated by the relaxation of
temperature and velocity differences between the constituent gases.
These relaxations appear to be captured adequately by the mixture
model of~\cite{Bonatto2005}.  This is the first time that these
predictions are tested in the kinetic regime.  The model contains no
adjustable parameters, and reproduces all experiments excellently.
The present successful comparison of model and experiment marks a step towards a
description of light scattering in mixtures of molecular gases (such as air),
and emphasizes the need to account for interspecies relaxation of
temperatures of both translational and kinetic degrees of freedom.


\begin{thebibliography}{28}
\expandafter\ifx\csname natexlab\endcsname\relax\def\natexlab#1{#1}\fi
\expandafter\ifx\csname bibnamefont\endcsname\relax
  \def\bibnamefont#1{#1}\fi
\expandafter\ifx\csname bibfnamefont\endcsname\relax
  \def\bibfnamefont#1{#1}\fi
\expandafter\ifx\csname citenamefont\endcsname\relax
  \def\citenamefont#1{#1}\fi
\expandafter\ifx\csname url\endcsname\relax
  \def\url#1{\texttt{#1}}\fi
\expandafter\ifx\csname urlprefix\endcsname\relax\def\urlprefix{URL }\fi
\providecommand{\bibinfo}[2]{#2}
\providecommand{\eprint}[2][]{\url{#2}}

\bibitem[{\citenamefont{{Strutt (Lord Rayleigh)}}(1899)}]{Strutt1899}
\bibinfo{author}{\bibfnamefont{J.~W.} \bibnamefont{{Strutt (Lord Rayleigh)}}},
  \bibinfo{journal}{Philos. Mag.} \textbf{\bibinfo{volume}{47}},
  \bibinfo{pages}{375} (\bibinfo{year}{1899}).

\bibitem[{\citenamefont{Brillouin}(1922)}]{Brillouin1922}
\bibinfo{author}{\bibfnamefont{L.}~\bibnamefont{Brillouin}},
  \bibinfo{journal}{Ann. d. Phys. (Paris)} \textbf{\bibinfo{volume}{17}},
  \bibinfo{pages}{88} (\bibinfo{year}{1922}).

\bibitem[{\citenamefont{Mandelstam}(1926)}]{Mandelstam1926}
\bibinfo{author}{\bibfnamefont{L.~I.} \bibnamefont{Mandelstam}},
  \bibinfo{journal}{Zh. Russ. Fiz-Khim.} \textbf{\bibinfo{volume}{58}},
  \bibinfo{pages}{381} (\bibinfo{year}{1926}).

\bibitem[{\citenamefont{{van Leeuwen} and Yip}(1965)}]{Leeuwen1965}
\bibinfo{author}{\bibfnamefont{J.~M.~J.} \bibnamefont{{van Leeuwen}}}
  \bibnamefont{and} \bibinfo{author}{\bibfnamefont{S.}~\bibnamefont{Yip}},
  \bibinfo{journal}{Phys. Rev. A} \textbf{\bibinfo{volume}{139}},
  \bibinfo{pages}{1138} (\bibinfo{year}{1965}).

\bibitem[{\citenamefont{She et~al.}(1992)\citenamefont{She, Alvarez, Caldwell,
  and Krueger}}]{She1992}
\bibinfo{author}{\bibfnamefont{C.~Y.} \bibnamefont{She}},
  \bibinfo{author}{\bibfnamefont{R.~J.} \bibnamefont{Alvarez}},
  \bibinfo{author}{\bibfnamefont{L.~M.} \bibnamefont{Caldwell}},
  \bibnamefont{and} \bibinfo{author}{\bibfnamefont{D.~A.}
  \bibnamefont{Krueger}}, \bibinfo{journal}{Opt. Lett.}
  \textbf{\bibinfo{volume}{17}}, \bibinfo{pages}{541} (\bibinfo{year}{1992}).

\bibitem[{\citenamefont{Witschas et~al.}(2014)\citenamefont{Witschas, Lemmerz,
  and Reitebuch}}]{Witschas2014b}
\bibinfo{author}{\bibfnamefont{B.}~\bibnamefont{Witschas}},
  \bibinfo{author}{\bibfnamefont{C.}~\bibnamefont{Lemmerz}}, \bibnamefont{and}
  \bibinfo{author}{\bibfnamefont{O.}~\bibnamefont{Reitebuch}},
  \bibinfo{journal}{Opt. Lett.} \textbf{\bibinfo{volume}{39}},
  \bibinfo{pages}{1972} (\bibinfo{year}{2014}).

\bibitem[{\citenamefont{Reitebuch et~al.}(2009)\citenamefont{Reitebuch,
  Lemmerz, Nagel, Paffrath, Durand, Endemann, Fabre, and
  Chaloupy}}]{Reitebuch2009}
\bibinfo{author}{\bibfnamefont{O.}~\bibnamefont{Reitebuch}},
  \bibinfo{author}{\bibfnamefont{C.}~\bibnamefont{Lemmerz}},
  \bibinfo{author}{\bibfnamefont{E.}~\bibnamefont{Nagel}},
  \bibinfo{author}{\bibfnamefont{U.}~\bibnamefont{Paffrath}},
  \bibinfo{author}{\bibfnamefont{Y.}~\bibnamefont{Durand}},
  \bibinfo{author}{\bibfnamefont{M.}~\bibnamefont{Endemann}},
  \bibinfo{author}{\bibfnamefont{F.}~\bibnamefont{Fabre}}, \bibnamefont{and}
  \bibinfo{author}{\bibfnamefont{M.}~\bibnamefont{Chaloupy}},
  \bibinfo{journal}{J. Atmos. Oceanic Technol.} \textbf{\bibinfo{volume}{26}},
  \bibinfo{pages}{2501} (\bibinfo{year}{2009}).

\bibitem[{\citenamefont{Boley et~al.}(1972)\citenamefont{Boley, Desai, and
  Tenti}}]{Boley1972}
\bibinfo{author}{\bibfnamefont{C.~D.} \bibnamefont{Boley}},
  \bibinfo{author}{\bibfnamefont{R.~C.} \bibnamefont{Desai}}, \bibnamefont{and}
  \bibinfo{author}{\bibfnamefont{G.}~\bibnamefont{Tenti}},
  \bibinfo{journal}{Can. J. Phys.} \textbf{\bibinfo{volume}{50}},
  \bibinfo{pages}{2158} (\bibinfo{year}{1972}).

\bibitem[{\citenamefont{Tenti et~al.}(1974)\citenamefont{Tenti, Boley, and
  Desai}}]{Tenti1974}
\bibinfo{author}{\bibfnamefont{G.}~\bibnamefont{Tenti}},
  \bibinfo{author}{\bibfnamefont{C.~D.} \bibnamefont{Boley}}, \bibnamefont{and}
  \bibinfo{author}{\bibfnamefont{R.~C.} \bibnamefont{Desai}},
  \bibinfo{journal}{Can. J. Phys.} \textbf{\bibinfo{volume}{52}},
  \bibinfo{pages}{285} (\bibinfo{year}{1974}).

\bibitem[{\citenamefont{Grinstead and Barker}(2000)}]{Grinstead2000}
\bibinfo{author}{\bibfnamefont{J.~H.} \bibnamefont{Grinstead}}
  \bibnamefont{and} \bibinfo{author}{\bibfnamefont{P.~F.}
  \bibnamefont{Barker}}, \bibinfo{journal}{Phys. Rev. Lett.}
  \textbf{\bibinfo{volume}{85}}, \bibinfo{pages}{1222 } (\bibinfo{year}{2000}).

\bibitem[{\citenamefont{Pan et~al.}(2002)\citenamefont{Pan, Shneider, and
  Miles}}]{Pan2002}
\bibinfo{author}{\bibfnamefont{X.}~\bibnamefont{Pan}},
  \bibinfo{author}{\bibfnamefont{M.~N.} \bibnamefont{Shneider}},
  \bibnamefont{and} \bibinfo{author}{\bibfnamefont{R.~B.} \bibnamefont{Miles}},
  \bibinfo{journal}{Phys. Rev. Lett.} \textbf{\bibinfo{volume}{89}},
  \bibinfo{pages}{183001} (\bibinfo{year}{2002}).

\bibitem[{\citenamefont{Pan et~al.}(2005)\citenamefont{Pan, Shneider, and
  Miles}}]{Pan2005}
\bibinfo{author}{\bibfnamefont{X.}~\bibnamefont{Pan}},
  \bibinfo{author}{\bibfnamefont{M.~N.} \bibnamefont{Shneider}},
  \bibnamefont{and} \bibinfo{author}{\bibfnamefont{R.~B.} \bibnamefont{Miles}},
  \bibinfo{journal}{Phys. Rev. A.} \textbf{\bibinfo{volume}{71}},
  \bibinfo{pages}{045801} (\bibinfo{year}{2005}).

\bibitem[{\citenamefont{Meijer et~al.}(2010)\citenamefont{Meijer, de~Wijn,
  Peters, Dam, and {W. van de Water}}}]{Meijer2010}
\bibinfo{author}{\bibfnamefont{A.~S.} \bibnamefont{Meijer}},
  \bibinfo{author}{\bibfnamefont{A.~S.} \bibnamefont{de~Wijn}},
  \bibinfo{author}{\bibfnamefont{M.~F.~E.} \bibnamefont{Peters}},
  \bibinfo{author}{\bibfnamefont{N.~J.} \bibnamefont{Dam}}, \bibnamefont{and}
  \bibinfo{author}{\bibnamefont{{W. van de Water}}}, \bibinfo{journal}{J. Chem.
  Phys.} \textbf{\bibinfo{volume}{133}}, \bibinfo{pages}{164315}
  (\bibinfo{year}{2010}).

\bibitem[{\citenamefont{Vieitez et~al.}(2010)\citenamefont{Vieitez, van Duijn,
  Ubachs, Witschas, Meijer, {de Wijn}, Dam, and {van de Water}}}]{Vieitez2010}
\bibinfo{author}{\bibfnamefont{M.~O.} \bibnamefont{Vieitez}},
  \bibinfo{author}{\bibfnamefont{E.~J.} \bibnamefont{van Duijn}},
  \bibinfo{author}{\bibfnamefont{W.}~\bibnamefont{Ubachs}},
  \bibinfo{author}{\bibfnamefont{B.}~\bibnamefont{Witschas}},
  \bibinfo{author}{\bibfnamefont{A.}~\bibnamefont{Meijer}},
  \bibinfo{author}{\bibfnamefont{A.~S.} \bibnamefont{{de Wijn}}},
  \bibinfo{author}{\bibfnamefont{N.~J.} \bibnamefont{Dam}}, \bibnamefont{and}
  \bibinfo{author}{\bibfnamefont{W.}~\bibnamefont{{van de Water}}},
  \bibinfo{journal}{Phys. Rev. A} \textbf{\bibinfo{volume}{82}},
  \bibinfo{pages}{043836} (\bibinfo{year}{2010}).

\bibitem[{\citenamefont{Gu and Ubachs}(2013)}]{Gu2013b}
\bibinfo{author}{\bibfnamefont{Z.}~\bibnamefont{Gu}} \bibnamefont{and}
  \bibinfo{author}{\bibfnamefont{W.}~\bibnamefont{Ubachs}},
  \bibinfo{journal}{Opt. Lett.} \textbf{\bibinfo{volume}{38}},
  \bibinfo{pages}{1110} (\bibinfo{year}{2013}).

\bibitem[{\citenamefont{Gu et~al.}(2014)\citenamefont{Gu, Ubachs, and van~de
  Water}}]{Gu2014a}
\bibinfo{author}{\bibfnamefont{Z.}~\bibnamefont{Gu}},
  \bibinfo{author}{\bibfnamefont{W.}~\bibnamefont{Ubachs}}, \bibnamefont{and}
  \bibinfo{author}{\bibfnamefont{W.}~\bibnamefont{van~de Water}},
  \bibinfo{journal}{Opt. Lett.} \textbf{\bibinfo{volume}{39}},
  \bibinfo{pages}{3301} (\bibinfo{year}{2014}).

\bibitem[{\citenamefont{Gu et~al.}(2013)\citenamefont{Gu, Witschas, van~de
  Water, and Ubachs}}]{Gu2013a}
\bibinfo{author}{\bibfnamefont{Z.}~\bibnamefont{Gu}},
  \bibinfo{author}{\bibfnamefont{B.}~\bibnamefont{Witschas}},
  \bibinfo{author}{\bibfnamefont{W.}~\bibnamefont{van~de Water}},
  \bibnamefont{and} \bibinfo{author}{\bibfnamefont{W.}~\bibnamefont{Ubachs}},
  \bibinfo{journal}{Appl. Opt.} \textbf{\bibinfo{volume}{52}},
  \bibinfo{pages}{4640} (\bibinfo{year}{2013}).

\bibitem[{\citenamefont{Gu and Ubachs}(2014)}]{Gu2014b}
\bibinfo{author}{\bibfnamefont{Z.~Y.} \bibnamefont{Gu}} \bibnamefont{and}
  \bibinfo{author}{\bibfnamefont{W.}~\bibnamefont{Ubachs}},
  \bibinfo{journal}{J. Chem. Phys.} \textbf{\bibinfo{volume}{141}},
  \bibinfo{pages}{104320} (\bibinfo{year}{2014}).

\bibitem[{\citenamefont{Gu et~al.}(2012)\citenamefont{Gu, Vieitez, van Duijn,
  and Ubachs}}]{Gu2012rsi}
\bibinfo{author}{\bibfnamefont{Z.}~\bibnamefont{Gu}},
  \bibinfo{author}{\bibfnamefont{M.~O.} \bibnamefont{Vieitez}},
  \bibinfo{author}{\bibfnamefont{E.~J.} \bibnamefont{van Duijn}},
  \bibnamefont{and} \bibinfo{author}{\bibfnamefont{W.}~\bibnamefont{Ubachs}},
  \bibinfo{journal}{Rev. Scient. Instrum.} \textbf{\bibinfo{volume}{83}},
  \bibinfo{pages}{053112} (\bibinfo{year}{2012}).

\bibitem[{\citenamefont{Chapman and Cowling}(1970)}]{Chapman1970}
\bibinfo{author}{\bibfnamefont{A.}~\bibnamefont{Chapman}} \bibnamefont{and}
  \bibinfo{author}{\bibfnamefont{T.~G.} \bibnamefont{Cowling}},
  \emph{\bibinfo{title}{Mathematical theory of non-uniform gases}}
  (\bibinfo{publisher}{Cambridge Mathematical library}, \bibinfo{year}{1970}),
  \bibinfo{edition}{3rd} ed., ISBN \bibinfo{isbn}{052140844}.

\bibitem[{\citenamefont{Bhatnagar et~al.}(1954)\citenamefont{Bhatnagar, Gross,
  and Krook}}]{Bhatnagar1954}
\bibinfo{author}{\bibfnamefont{P.~L.} \bibnamefont{Bhatnagar}},
  \bibinfo{author}{\bibfnamefont{E.~P.} \bibnamefont{Gross}}, \bibnamefont{and}
  \bibinfo{author}{\bibfnamefont{M.}~\bibnamefont{Krook}},
  \bibinfo{journal}{Phys. Rev.} \textbf{\bibinfo{volume}{94}},
  \bibinfo{pages}{511} (\bibinfo{year}{1954}).

\bibitem[{\citenamefont{{Marques Jr} and Kremer}(1998)}]{Marques1998}
\bibinfo{author}{\bibfnamefont{W.}~\bibnamefont{{Marques Jr}}}
  \bibnamefont{and} \bibinfo{author}{\bibfnamefont{G.~M.}
  \bibnamefont{Kremer}}, \bibinfo{journal}{Continuum Mech. Thermodyn.}
  \textbf{\bibinfo{volume}{10}}, \bibinfo{pages}{319} (\bibinfo{year}{1998}).

\bibitem[{\citenamefont{Bonatto and {Marques Jr.}}(2005)}]{Bonatto2005}
\bibinfo{author}{\bibfnamefont{J.~R.} \bibnamefont{Bonatto}} \bibnamefont{and}
  \bibinfo{author}{\bibfnamefont{W.}~\bibnamefont{{Marques Jr.}}},
  \bibinfo{journal}{J. Stat. Mech.} \textbf{\bibinfo{volume}{P09014}}
  (\bibinfo{year}{2005}).

\bibitem[{\citenamefont{Fernandes and {Marques Jr.}}(2004)}]{Fernandes2004}
\bibinfo{author}{\bibfnamefont{A.~S.} \bibnamefont{Fernandes}}
  \bibnamefont{and} \bibinfo{author}{\bibfnamefont{W.}~\bibnamefont{{Marques
  Jr.}}}, \bibinfo{journal}{Physica A} \textbf{\bibinfo{volume}{332}},
  \bibinfo{pages}{29} (\bibinfo{year}{2004}).

\bibitem[{\citenamefont{Clark}(1975)}]{Clark1975}
\bibinfo{author}{\bibfnamefont{N.~A.} \bibnamefont{Clark}},
  \bibinfo{journal}{Phys. Rev. A.} \textbf{\bibinfo{volume}{12}},
  \bibinfo{pages}{2092} (\bibinfo{year}{1975}).

\bibitem[{\citenamefont{Letamendia et~al.}(1982)\citenamefont{Letamendia,
  Joubert, Chabrat, Rouch, Vaucamps, Boley, Yip, and Chen}}]{Letamendia1982}
\bibinfo{author}{\bibfnamefont{L.}~\bibnamefont{Letamendia}},
  \bibinfo{author}{\bibfnamefont{P.}~\bibnamefont{Joubert}},
  \bibinfo{author}{\bibfnamefont{J.~P.} \bibnamefont{Chabrat}},
  \bibinfo{author}{\bibfnamefont{J.}~\bibnamefont{Rouch}},
  \bibinfo{author}{\bibfnamefont{C.}~\bibnamefont{Vaucamps}},
  \bibinfo{author}{\bibfnamefont{C.~D.} \bibnamefont{Boley}},
  \bibinfo{author}{\bibfnamefont{S.}~\bibnamefont{Yip}}, \bibnamefont{and}
  \bibinfo{author}{\bibfnamefont{S.~H.} \bibnamefont{Chen}},
  \bibinfo{journal}{Phys. Rev. A} \textbf{\bibinfo{volume}{25}},
  \bibinfo{pages}{481} (\bibinfo{year}{1982}).

\bibitem[{\citenamefont{Boley and Yip}(1972)}]{Boley1972b}
\bibinfo{author}{\bibfnamefont{C.~D.} \bibnamefont{Boley}} \bibnamefont{and}
  \bibinfo{author}{\bibfnamefont{S.}~\bibnamefont{Yip}},
  \bibinfo{journal}{Phys. of Plasmas} \textbf{\bibinfo{volume}{15}},
  \bibinfo{pages}{1424} (\bibinfo{year}{1972}).

\bibitem[{\citenamefont{Dicke}(1953)}]{Dicke1953}
\bibinfo{author}{\bibfnamefont{R.~H.} \bibnamefont{Dicke}},
  \bibinfo{journal}{Phys. Rev.} \textbf{\bibinfo{volume}{89}},
  \bibinfo{pages}{472} (\bibinfo{year}{1953}).

\end{thebibliography}

\end{document}